\documentclass[a4paper,10pt,english,pra, twocolumn, floatfix, showpacs]{revtex4-1}

\usepackage{amsmath}
\usepackage{amssymb}
\usepackage{mathtools}
\usepackage{graphicx}
\usepackage{hyperref}
\usepackage{color}

\newcommand{\figref}[1]{Fig.~\ref{#1}}

\begin{document}

\title{Self-Energy Dispersion in the Hubbard Model}
\author{Thomas Mertz}
\author{Karim Zantout}
\author{Roser Valent{\'{i}}}
\affiliation{Institut f\"ur Theoretische Physik, Goethe-Universit\"at, Frankfurt am Main, Germany}

\begin{abstract}
We introduce the concept of self-energy dispersion as an error bound on local theories and apply it to the two-dimensional Hubbard model on the square lattice at half-filling. Since the self-energy has no single-particle analog and is not directly measurable in experiments, its general behavior as a function of momentum is an open question. In this article we benchmark the momentum dependence with the Two-Particle Self-Consistent approach together with analytical and numerical considerations and we show that through the addition of a local single-particle potential to the Hubbard model the self-energy can be flattened, such that it is essentially described by only a frequency-dependent term. We use this observation to motivate that local theories, such as the dynamical mean-field theory, should be expected to give very accurate results in the presence of a potential of this kind. Finally, we propose a simple energy argument as an estimator for the crossover from non-local to local self-energies, which can be computed even by local theories such as dynamical mean-field theory.
\end{abstract}

\maketitle

The Hubbard model \cite{Hubbard1963,Imada1998} is one of the most prominent models in condensed matter physics. It describes correlated electrons in arbitrary lattices and predicts physical phenomena such as Mott insulators, topological phases or superconductivity. Although the Hubbard model is restricted to a local interaction, an exact analytical solution is still not available for dimensions larger than one. In this context, intensive efforts have recently been devoted to provide numerical benchmarks of ground-state and excited-state properties of the two-dimensional Hubbard model on the square lattice \cite{LeBlanc2015}.
A very popular approach to many-body physics in condensed matter is the Green's function formalism based on many-body perturbation theory, where the interaction effect is encoded in the one-particle irreducible part, the so-called self-energy. In static mean-field theory the self-energy reduces to a constant, which acts like a chemical potential, while in dynamical mean-field theory (DMFT) \cite{Georges1996, Kotliar2004} the self-energy is a frequency dependent function, which can cause a redistribution of spectral weight. In this work, we are interested in the momentum dependence, which is not resolved in a local theory such as DMFT due to the omission of non-local diagrams \cite{Toschi2007}.

The effect of a momentum dependence varies between different physical observables. Integrated quantities, such as the spectral weight $A(\omega)$, are expected to be rather well-described by a local theory, while spectral functions $A(k,\omega)$, Fermi surfaces or structure factors, which are explicit measures of a system's dispersion, can be affected substantially by the momentum dependence compared to the mean-field case.

The approximations of DMFT become exact whenever the self-energy is purely local, as it is the case e.g.~in infinite dimensions \cite{Metzner1989}. Here we show that this case can also be realized by adding single-particle terms, such as an alternating potential, to the Hamiltonian \cite{Fabrizio1999,Bouadim2007,Paris2007,Kim2014,Hafez2016,Samanta2016,Miyao2016,Loida2017}. These terms are known to localize the system if the coupling is much stronger than the hopping. We will first discuss these limiting cases in the presence of a local Hubbard interaction. Within the framework of the Two-Particle Self-Consistent approach (TPSC) \cite{Vilk1997,Zantout2018}, where the momentum dependence of the self-energy is included in a non-perturbative way, we then identify the non-trivial regimes where the self-energy is still local with good accuracy and hence where DMFT is expected to yield very precise results as well.
For a direct comparison we perform additional DMFT and exact diagonalization (ED) calculations, which---for the purpose of comparability---are restricted to non-magnetic solutions. In addition, we will develop a physical argument to back up our numerical results.

\begin{figure}
	\includegraphics[width=0.8\linewidth]{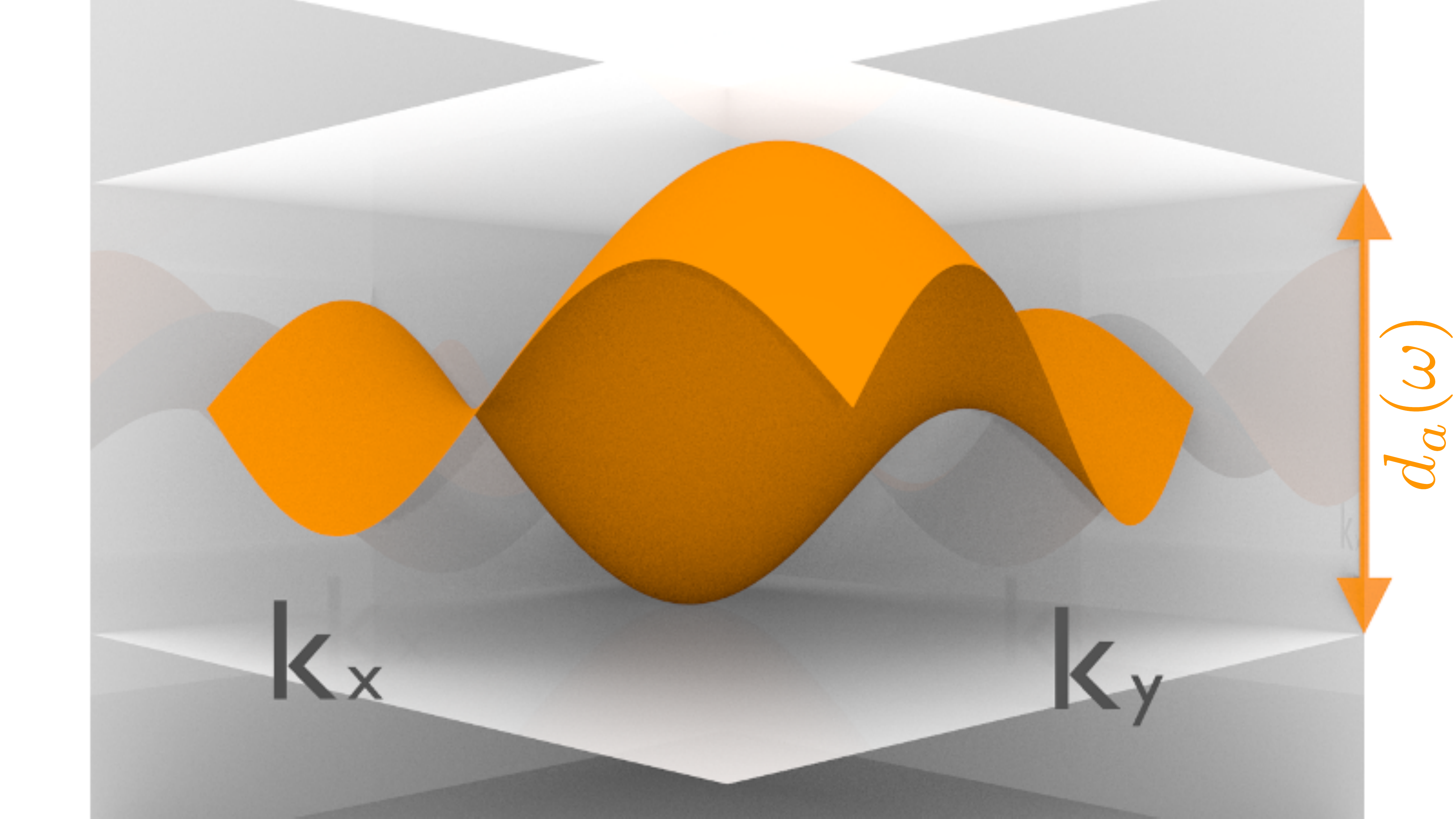}
	\caption{Illustration of the self-energy dispersion strength $d_a$ eq.~\eqref{eq:dispersion_strength} as the thickness of the self-energy surface over the Brillouin zone. Here, we show only one matrix element; typically we use the dispersion strength $d_a$, which is the maximal thickness of all the matrix elements.}
	\label{fig:sigma_surface}
\end{figure}

\textit{Self-Energy Dispersion Amplitude.}---In order to quantitatively describe the momentum-dependence or non-locality of the self-energy we define the absolute self-energy dispersion amplitude
\begin{equation}
	d_a(\omega) = \mathrm{max}_{k,k^\prime} \lVert\Sigma(k, \omega) - \Sigma(k^\prime, \omega)\rVert,
	\label{eq:dispersion_strength}
\end{equation}
at a fixed frequency $\omega$,
where the notation $\lVert \cdot\rVert$ denotes the maximum norm. From eq.~\eqref{eq:dispersion_strength} it is evident that $d_a$ is simply the \textit{thickness} of the self-energy surface at constant $\omega$, cf.~\figref{fig:sigma_surface}.
A better indication of the importance of the self-energy dispersion might be conveyed by the relative dispersion amplitude
\begin{equation}
	d_r(\omega) = \begin{cases}
	\frac{d_a(\omega)}{N_k^{-1}\lVert\sum_k \Sigma(k, \omega)\rVert} \;&\mathrm{if}~ \sum_k \Sigma(k, \omega)\neq 0\\
	d_a(\omega) \;&\mathrm{else},
	\end{cases}
	\label{eq:dispersion_strength_rel}
\end{equation}
which measures the variation of $\Sigma$ throughout the Brillouin zone compared to the average value, i.e.~the local self-energy.
Naturally, both $d_a$ and $d_r$ will vanish if $\Sigma(k, \omega)=0$.

The advantage of this definition is that the error of the DMFT approximation can be expressed in terms of the dispersion amplitude (cf.~Appendix \ref{app:dispersion})
\begin{equation}
	\varepsilon(\omega) = \lVert \Sigma_\mathrm{exact}(k, \omega) - \Sigma_\mathrm{DMFT}(\omega)\rVert 
	\leq d_a(\omega) + r(\omega),
	\label{eq:dmft_error}
\end{equation}
where the positive function $r(\omega)$ is an additional error of the local self-energy in DMFT due to the approximate nature of the self-consistency cycle in finite dimensions, which vanishes if the exact self-energy is local.
Moreover, $d_a$ and $r$ are not independent functions, since in the local limit, where DMFT becomes exact, both functions must vanish. One can show by contradiction that $r$ cannot be finite while $d_a$ vanishes, since this would imply non-locality. Therefore, we propose that in most cases $d_a$ should be a good indicator to judge the validity of the local approximation, since small $d_a$ implies small $r$.

\begin{figure}[b]
	\includegraphics[width=\linewidth]{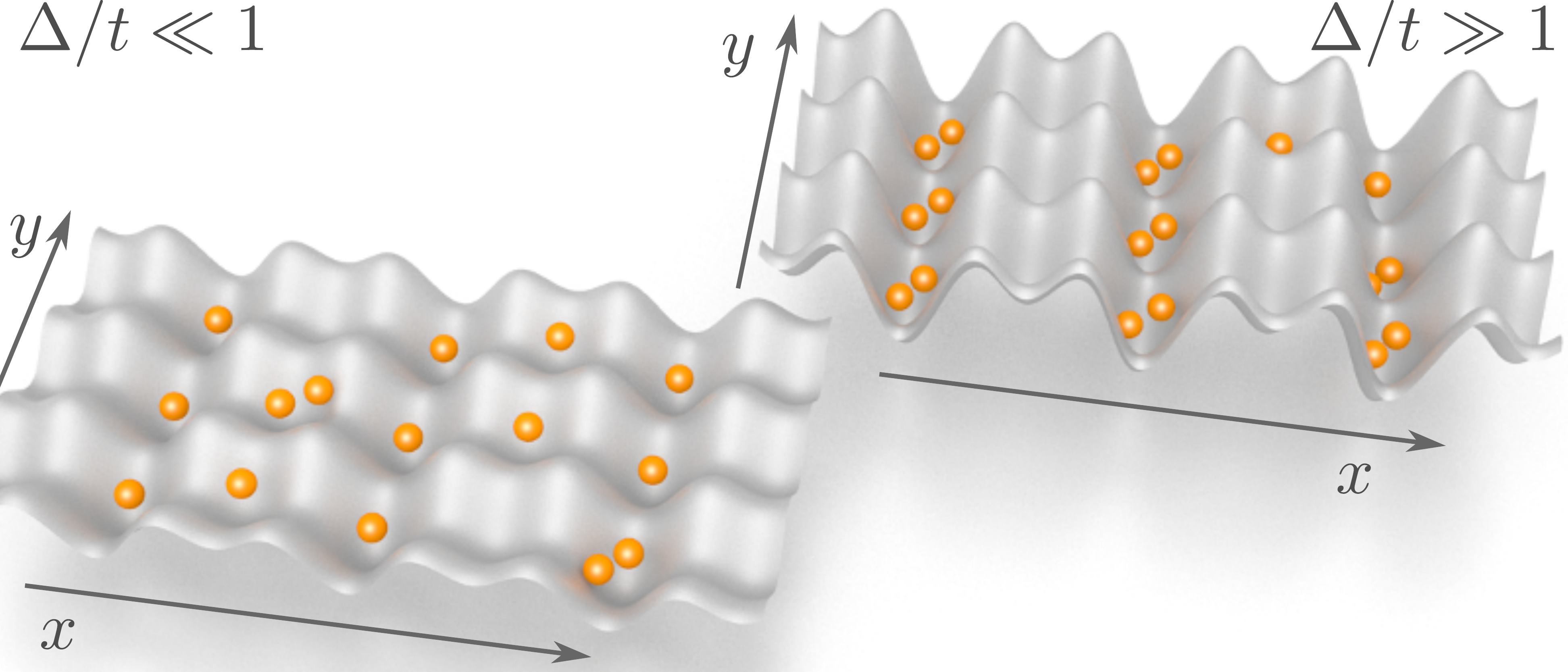}
	\caption{Illustration of the electron density in a shallow and deep on-site potential with amplitude $\Delta/t$, cf.~eq.~\eqref{eq:ionic_hubbard_model}. For small $\Delta/t$ (lower left) the average densities are near unity on $A$ and $B$ sites. For large $\Delta/t$ (upper right) $A$ sites will be filled almost completely while $B$ sites are almost empty.}
	\label{fig:staggering}
\end{figure}

\textit{Ionic Hubbard Model.}---In the following we study the ionic Hubbard model at half filling
\begin{equation}
\begin{split}
	H &= -t\sum_{\langle i, j\rangle} c_i^\dagger c_j - \Delta \sum_{i\in A} n_i + \Delta \sum_{i\in B} n_i + U \sum_i n_{i\uparrow} n_{i\downarrow}.
\end{split}
\label{eq:ionic_hubbard_model}
\end{equation}
Here, $c_i, c_i^\dagger$ are the fermionic operators on site $i$ (an additional spin index is implied) and $\Delta\geq 0$ is the amplitude of the ionic potential. We split the square lattice into two sublattices (columns labelled by $A,B$) with alternating potential along the $x$-direction, which produces a variant of the ionic Hubbard model \cite{Paris2007}. In the usual formulation of the ionic Hubbard model the potential alternates in both directions, which freezes out the system much more quickly and is therefore less interesting for our study. Sites belonging to the $A$ sublattice will have a local chemical potential of $-\Delta$ and vice versa for $B$ sites.
As a result, the $A$ sublattice will show a higher occupation, i.e.~charge order is introduced. Due to the energetically unfavorable double occupation of any site, the Hubbard interaction tends to produce many-body ground states with average occupations more uniformly spread among the sites of both sub-lattices, which is exactly opposite to the effect of the local potential. Therefore, one could expect that the interacting system is far less local than the non-interacting system as $U$ partly compensates for the effect of $\Delta$ and vice versa, as seen in e.g.~the delay of the antiferromagnetic transition on the 2D square lattice \cite{Bag2015}. Typical configurations for systems subject to weak and strong $\Delta$ potentials are illustrated in \figref{fig:staggering}.

\textit{Limiting Cases.}---There are two simple limits for this model. Let us first investigate $U\rightarrow\infty$ and $\frac{U}{\Delta}>1$. the system orders magnetically (AF order in the square lattice), which leads to a suppressed double occupancy on both sublattices. We note that this suppressed local density-density correlation is independent of the magnetic properties of the system, i.e.~it is also observed in systems which exhibit magnetic frustration. This also leads to a momentum-independent self-energy as the dynamics in the lattice freeze out due to the strong repulsion.
The self-energy in the atomic limit is dispersion-less, as one can see from the Green's function on the $A$ lattice in the ground state, which is given by
\begin{equation}
	G_{AA}(k, \omega) = \frac{\omega + \Delta}{(\omega+\Delta)^2- \frac{U^2}{4}}, \quad \Sigma_{AA}(k, \omega) = \frac{U^2}{4(\omega+\Delta)}.
\end{equation}
In the opposite case, where $\Delta\rightarrow\infty$, the single particle energy of states localized on sublattice $B$ is infinite, i.e.~the entire sublattice is essentially vacant. Due to the half filling condition and the fact that the system contains equally many $A$ and $B$ sites, this means that all $A$ sites must be doubly occupied. At finite $U$ this is energetically favored over partial occupation of $B$ sites, such that $\langle n_{A\uparrow} n_{A\downarrow}\rangle = 1$, $\langle n_{B\uparrow} n_{B\downarrow}\rangle = 0$.
Assuming that $\frac{U}{\Delta}<1$ we obtain the Green's function and self-energy
\begin{equation}
	G_{AA}(k, \omega) = \left[\omega - \frac{U}{2} + \Delta\right]^{-1}, \quad
	\Sigma_{AA}(k, \omega) = \frac{U}{2},
\end{equation}
respectively, on the $A$ sublattice. The self-energy is a purely local function, since exchange between sites is strongly suppressed, cf.~Appendix~\ref{app:local_selfenergy}.

The above expressions are strict limits and it is unclear how close to the atomic limit the system becomes increasingly local.
Our numerical data, which we will present in the following, suggests that in the intermediate region, where $U\gg t$, the dispersion of the self-energy will increase in significance, which results in increasing values of the relative dispersion strength $d_r$.

\textit{Results.}---We first study the usual Hubbard model on a square lattice, where $\Delta=0$.
The square lattice is known to show very strong correlations manifested in an antiferromagnetic instability at low temperatures, which is why we expect it to be a limiting case for our study. The strong correlations result also in a strong momentum-dependence of the self-energy at low temperatures. All our calculations are done at an inverse temperature of $\beta t=10$, which is low enough for us to observe the pseudo-gap phase \cite{Vilk1997}, where the spectral density is reduced at the Brillouin zone boundaries, leading to an especially strong momentum-dependence. We also checked that the features we observe in the Green's function at finite temperature correspond to those observed at zero temperature in cluster perturbation theory (CPT) \cite{Gros1993,Senechal2002}. The momentum-dependent TPSC self-energy for this model is also in agreement with the dual-fermion result obtained in Refs.~\cite{Rubtsov2008,Rubtsov2009}.

\begin{figure}[h]
	\includegraphics[scale=1]{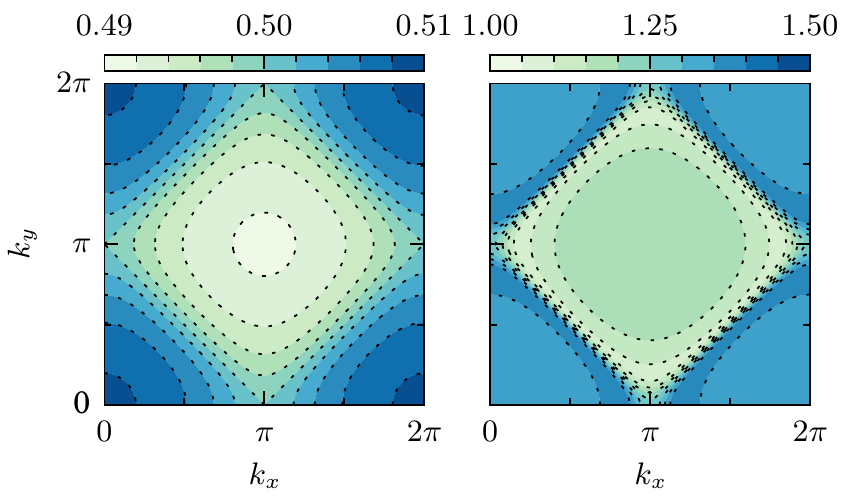}
\caption{Self-energy $\mathrm{Re}\Sigma_\mathrm{TPSC}(k, i\omega_{n=0})$ of the square lattice at frequency $i\omega_0$, $\beta t=10$, at $U/t=1$ (left) and $U/t=2.5$ (right). The self-energy has a dispersion around the Hartree-Fock value $U/2$, which is very weak at low $U$ and increases with $U$.}
\label{fig:sigma_comp}
\end{figure}
\begin{figure}
	\includegraphics[scale=1]{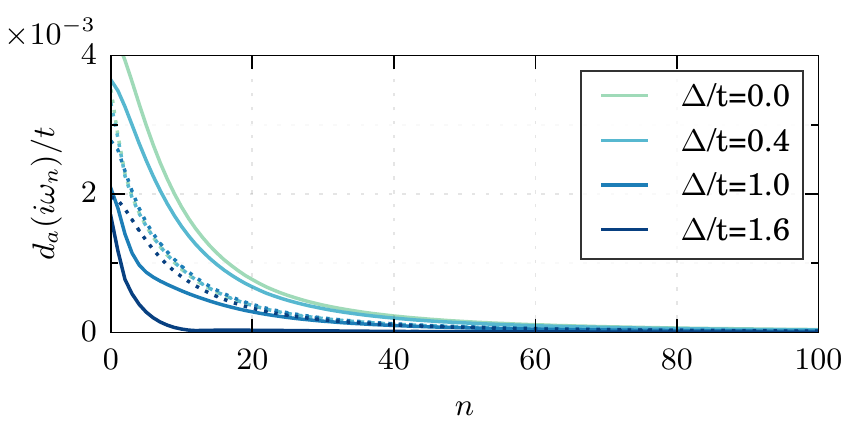}
	\caption{Frequency-resolved self-energy dispersion strength $d_a(i\omega_n)$ for diagonal matrix elements (lines) and off-diagonal matrix elements (dotted) at $U/t=0.6$. The dispersion strength decays rather quickly towards higher Matsubara frequencies and is maximal at $\omega_{n=0}=\pi/\beta$. Off-diagonal elements are only weakly dependent on $\Delta$} %
	\label{fig:dispersion_vs_frequency}
\end{figure}

In \figref{fig:sigma_comp} we compare the self-energies $\Sigma_\mathrm{TPSC}(k, i\omega_{n=0})$ at weak and intermediate coupling. We observe that the weak momentum dependence at $U/t=1$ becomes very substantial at intermediate interaction strengths $U/t=2.5$. Due to the power law decay of the self-energy we chose here to investigate the smallest frequency $i\omega_{n=0}$, which is expected to show the strongest dispersion as demonstrated in \figref{fig:dispersion_vs_frequency} ($\Delta/t=0$ line).
Altough TPSC is non-perturbative, we are bound to study the low- to intermediate-$U$ regime as a consequence of the limited validity of the approximation at larger $U$.

Having benchmarked the results of TPSC at $\Delta=0$, we now study the effect of a finite $\Delta$ potential on the momentum-dependence of the self-energy in the Hubbard model.
From \figref{fig:dispersion_vs_frequency} it is evident that also at $\Delta>0$ the self-energy dispersion decays rapidly as a function of frequency, which enables us to extract an upper bound on the error of the local approximation from the zero-frequency self-energy dispersion strength $d_a(i\omega_0)$. Our data also reveals that the dispersion of the off-diagonal matrix elements decays much slower as a function of $\Delta$.

\begin{figure}[h]
	\includegraphics[scale=1]{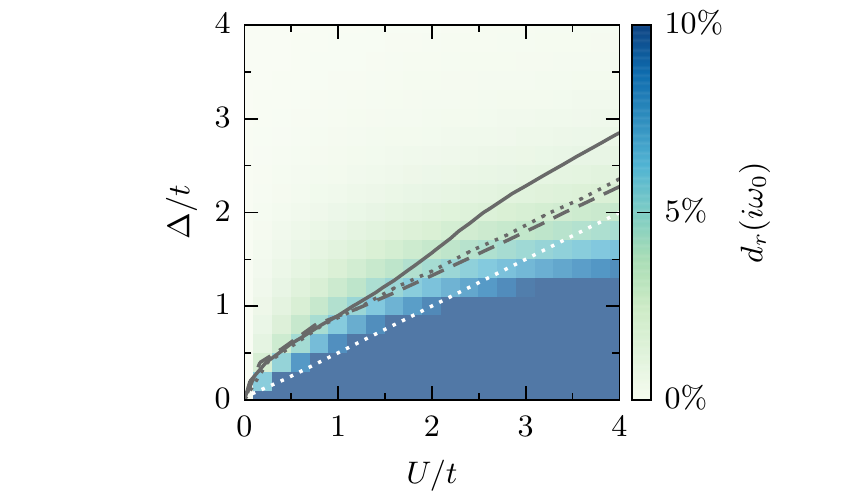}
	\caption{Relative dispersion strength $d_r(i\omega_{n=0})$ for the square lattice at inverse temperature $\beta t=10$, for better visibility of small dispersion amplitudes we have fixed the upper limit of the color scale to 0.1. In addition we plot the critical estimation of eq.~\eqref{eq:critical_delta} obtained for $t=0$ (white line) and in grey: TPSC (solid), ED (dashed), DMFT (dotted).}
	\label{fig:square_dispersion_strength}
\end{figure}

In \figref{fig:square_dispersion_strength} we plot (in color) the relative dispersion strength of the self-energy $d_r(\omega_{0})$, cf.~eq.~\eqref{eq:dispersion_strength_rel}, computed with TPSC over a range from low to intermediate $\Delta$ and $U$. We confirmed that these results can also be reproduced qualitatively at lower temperatures $\beta t=20$. 
The aforementioned effect of a diminished self-energy dispersion as a consequence of the alternating potential is clearly observed here. Compared to the Hartree value, the dispersion varies only within a few percent. We find that above $\Delta/t\approx 2$ the dispersion virtually vanishes everywhere in the regime studied. 

Apparently, there is a transition from strong momentum-dependence to a momentum-independent self-energy as a function of $\Delta$.
In order to give an estimate for the transition we propose the following analysis. The transition from strong to weak dispersion at weak to intermediate coupling is driven by the competition between $U$ and $\Delta$. The interaction $U$ induces a momentum-dependence via non-local diagrams and the potential $\Delta$ flattens everything out as it acts like a strong shielding term. In order to find the point where none of the two terms outweighs the other, we compare the energy cost corresponding to each potential
\begin{equation}
	2\Delta_c = U_c,
	\label{eq:critical_delta_hartree}
\end{equation}
where $2\Delta$ is the energy difference between $A$ and $B$ sites, which has to be paid for a homogeneous distribution of the fermions, and $U$ is the excitation energy needed to doubly occupy a single site, which is preferred by the $\Delta$ potential. Eq.~\eqref{eq:critical_delta_hartree} can also be derived in the mean-field approximation at large $\Delta/U$.

Beyond mean field we expect the crossover from strongly to weakly momentum-dependent self-energies to happen when the energy expectation values corresponding to the Hubbard repulsion and single particle potential, respectively, are equal, which yields
\begin{equation}
	\Delta_c = U_c \frac{\langle n_\uparrow^A n_\downarrow^A \rangle + \langle n_\uparrow^B n_\downarrow^B \rangle}{\langle n_A \rangle - \langle n_B \rangle} =\vcentcolon U_c \frac{D_A + D_B}{n_A - n_B}.
	\label{eq:critical_delta}
\end{equation}
$\Delta_c$ in eq.~\eqref{eq:critical_delta} is positive, since $\langle n_A\rangle \geq \langle n_B\rangle$ and the equal sign applies only when $\Delta/U=0$.
Since the numerator in eq.~\eqref{eq:critical_delta} is $2 \geq D_A + D_B \geq 0$ and the denominator satisfies $2 \geq n_A - n_B \geq 0$ we cannot obtain trivial bounds for the location of the transition.

In the atomic limit the zero temperature ground state is known exactly. For our system we obtain the double occupancy
\begin{equation}
	D_A = \Theta(2\Delta -U),
\end{equation}
which indicates a first-order phase transition from an antiferromagnet to a paramagnet at $U=2\Delta$. Incidentally, this line $\Delta(U) = U/2$ corresponds to the estimate of eq.~\eqref{eq:critical_delta_hartree} and eq.~\eqref{eq:critical_delta}, i.e.~at $t/U=0$ eq.~\eqref{eq:critical_delta_hartree} is the phase separation line of a first-order transition. At finite $t/U$ the transition is broadened into a crossover and we can only establish a typical region for the crossover to take place, which coincides with the region where the self-energy loses its momentum-dependence.
At finite $T$ and $t/U=0$ one obtains an exact solution to eq.~\eqref{eq:critical_delta}
\begin{equation}
	U_c = 2\Delta_c\tanh(\beta\Delta_c),
	\label{eq:transition_atomic}
\end{equation}
which has a broadened transition around the $T=0$ result $U=2\Delta$.

In \figref{fig:square_dispersion_strength} we show that this intuition provides a rather good description of the self-energy dispersion by comparing the critical line of eq.~\eqref{eq:critical_delta} computed with TPSC, DMFT and exact diagonalization of finite clusters (ED).
For the ED calculations we use the Lanczos procedure on various 8- and 12-site clusters for the same parameters at both $T=0$ and at $T=0.1 t$, in which $\Delta_c$ displayed only marginal cluster size- or temperature-dependence, and found good agreement with the TPSC prediction at low $\Delta/t,U/t$. At larger $U/t$, ED predicts that the transition line lies closer to the $t=0$ limit, which also agrees better with the self-energy dispersion amplitude.
For our DMFT calculations we use the continuous time hybridization expansion quantum Monte Carlo impurity solver from the ALPS package \cite{Bauer2011,Werner2006,Hafermann2013}.
The critical value $\Delta_c$ obtained with DMFT shows the same qualitative behavior as the data from TPSC and ED, which suggests that eq.~\eqref{eq:critical_delta} can be used to test the quality of the local approximation in DMFT. All calculations consistently produce a critical value larger than that obtained at $t=0$, since the additional kinetic term suppresses the splitting in the sublattice occupations.
\begin{figure}[h]
	\includegraphics[scale=1]{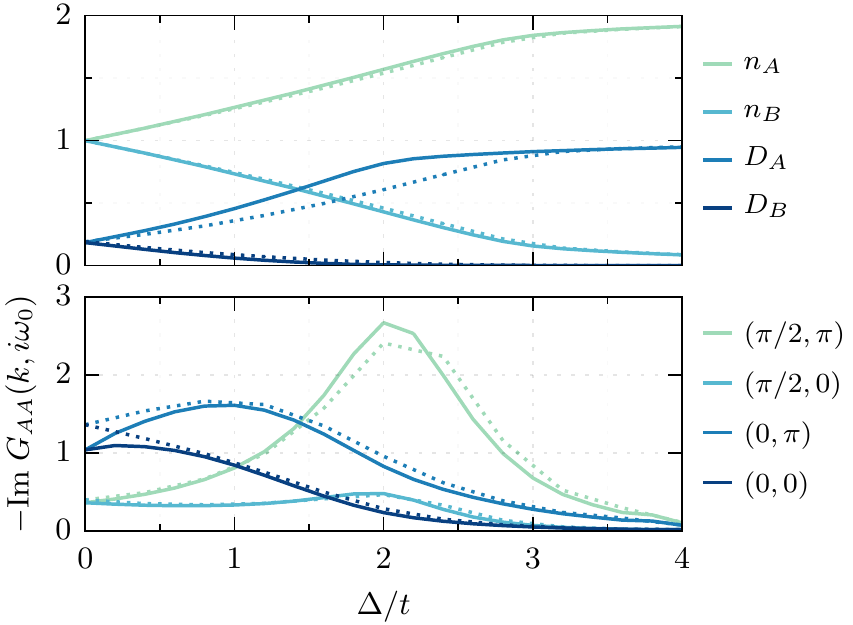}
	\caption{We compare TPSC (lines) with DMFT (dotted) at $U/t=2$. Top: Densities and double occupancies on $A$ and $B$ sites of the square lattice. Both methods agree at small and large $\Delta/t$, at moderate $\Delta/t$ the values for $D_A$ differ. Bottom: imaginary part of the Green's function in units of $t^{-1}$ for the $A$ site at the lowest Matsubara frequency at different momenta. At $(0, 0)$ and $(0, \pi)$ the momentum dependence leads to a reduced spectral weight, at $\Delta/t\gtrsim 1$ there is no significant difference.}
	\label{fig:locals}
\end{figure}

We now show---by comparing densities and Green's functions directly---that the self-energy dispersion strength also indicates an error in measurable quantities. In \figref{fig:locals} we compare the local observables density $n_{A,B}$ and double occupancy $D_{A,B}$ and the imaginary part of the Matsubara Green's function at the lowest frequency. The latter is a surrogate for the spectral function with the benefit of being free of an additional error due to analytic continuation.
We observe that the local observables $n_{A,B}$ and $D_{A,B}$ obtained from either TPSC or DMFT are essentially equal except for a deviation in the double occupancy which is a limitation of the ansatz in the solution to the sum rules underlying TPSC. The spectral function differs at low $\Delta/t$ at certain $k$-points, but quickly converges towards the DMFT value at $\Delta/t\approx 1$, which is in agreement with our observation of the self-energy becoming local at approximately this point, cf.~\figref{fig:square_dispersion_strength}.

\textit{Conclusion.}---In this article we have introduced the concept of the self-energy dispersion strength, which lends itself as an upper bound to the DMFT error.
We have carried out a numerical investigation of the momentum-dependence of the self-energy within the framework of TPSC for a variant of the ionic Hubbard model on the 2D square lattice, which we have benchmarked against additional calculations using DMFT and ED. Our findings include the suppression or even outright absence of a relevant self-energy dispersion in the presence of a sufficiently strong staggering field, which is expected in the static limit, where $t=0$. Based on a simple energy argument we have identified a characteristic measure, dependent on equal time one- and two-particle correlation functions, which let us determine the approximate position of the non-local to local crossover in the square lattice as a function of $U$ and $\Delta$. We have shown that in a large parameter regime the TPSC self-energy is rather dispersion-less and therefore previous DMFT studies of the ionic Hubbard model are expected to be very good and improvements using methods including non-local effects should be very moderate.
Due to the physical interpretation of an essentially fully occupied sublattice we assume that the present findings of a local limit should quite generally apply to any model on a lattice with an energy offset between two equal-sized sublattices.

\textit{Acknowledgments.}---T.~Mertz thanks D.~Kaib for useful discussions. R.~Valent{\'i} and K.~Zantout acknowledge financial support by the Deutsche Forschungsgemeinschaft through grant SFB/TR 49.

\bibliographystyle{ieeetr}
\bibliography{bibliography}

\appendix

\section{Proof of Eq.~\eqref{eq:dmft_error}}
\label{app:dispersion}
In eq.~\eqref{eq:dmft_error} we suggest that the DMFT error is bounded by a sum of the absolute self-energy dispersion amplitude $d_a(\omega)$ and a dependent positive function $r(\omega)$.
In order to show this, we expand the exact self-energy around the DMFT approximation, which is the principal idea behind the dual fermion approach \cite{Rubtsov2008,Rubtsov2009}. For this argument the expansion does not have to be systematic, i.e.~we can simply use the usual Feynman weak-coupling perturbation expansion and sum up all local diagrams to the DMFT self-energy. In general, we can express the corrections as a function $S(k, \omega)$ or more conveniently as
\begin{equation}
	\Sigma_\mathrm{exact}(k, \omega) = \Sigma_\mathrm{DMFT}(\omega) + S_0(k,\omega) + S_1(k, \omega),
\end{equation}
where $\sum_k S_0 = 0$ and $\sum_k S_1 \neq 0$. The partial sum $S_1$ therefore contains additional corrections to the local part, which can arise due to the self-consistency based on an invalid local approximation in DMFT.
We define $R(\omega) = \sum_k S_1(k, \omega)$ and note that $S_1 - R$ can now be accumulated into $S_0$ as well by defining $S_0^\prime=S_0+S_1-R$. Therefore, we have
\begin{equation}
	\Sigma_\mathrm{exact}(k, \omega) = \Sigma_\mathrm{DMFT}(\omega) + S^\prime_0(k,\omega) + R(\omega),
	\label{eq:sigma_exp}
\end{equation}
where the momentum-dependence is encoded entirely in a sum of diagrams $S_0^\prime$, which has a vanishing momentum-average. This property is important, as it allows us to identify
\begin{equation}
	\lVert S^\prime_0(k, \omega) \rVert = d_a(\omega).
\end{equation}
Hence, by defining $r(\omega) = \lVert R(\omega) \rVert$ the DMFT error therefore has an upper bound given by eq.~\eqref{eq:dmft_error}.

\section{Diagrammatic Explanation for the Local Self-Energy Limit}
\label{app:local_selfenergy}
Since the self-energy is first and foremost a manybody quantity it is not apparent why a local single-particle term affects the dispersion of the self-energy. Here we show that one can understand this from a simple Feynman diagram picture.

According to the standard Feynman rules any $n$th order term in the self-energy perturbation series can be expressed in terms of some Feynman diagram, which contains a number of $n$ interaction vertices and $2n$ propagators. We here assume that the diagram is not a skeleton diagram and therefore the propagators are all free propagators $G_0$. A momentum dependence occurs if the self-energy has contributions from diagrams containing vertices at different positions $i, j$, which are connected by propagators $G_0^{ij}$. If we assume that $i,j$ belong to different sublattices then $G_0^{ij}\rightarrow 0$. Therefore, only local diagrams containing $G_0^{ii}$ contribute.

\section{TPSC}
\label{app:tpsc}

In this work we compute the TPSC self-energy, which is defined as \cite{Zantout2018}
\begin{equation}
\begin{split}
	\Sigma_{ij,\sigma}(k) = U n^0_{i,\bar{\sigma}}\delta_{ij} + \frac{U}{8}\frac{T}{N} \sum_{q} \left[ 3\chi_{ij}^\mathrm{sp}(q)U_{jj}^\mathrm{sp}\right. \\
	\left. + \chi_{ij}^\mathrm{ch}(q)U_{jj}^\mathrm{ch}\right] G^0_{ij,\sigma}(k-q),
\end{split}
\end{equation}
where the superscript 0 indicates the non-interacting density and Green's function.
The self-energy is thus computed in a ``single shot" calculation, however, the two-particle vertices $U_\mathrm{ch}$ and $U_\mathrm{sp}$ and the spin and charge susceptibilities $\chi^\mathrm{sp}$ and $\chi^\mathrm{ch}$ are effectively determined self-consistently, as they depend on the self-consistent solution for the double occupancy $\langle n_\uparrow n_\downarrow\rangle$.
Note that the first term, which is essentially a constant with neither momentum nor frequency dependence, is exactly the Hartree diagram in first order perturbation theory. 
At various fillings and intermediate interactions it has been shown that TPSC agrees well with Quantum Monte Carlo calculations \cite{Vilk1994}.
However, in the present study, where an additional energy offset leads to a partition of the lattice, the Hartree term is no longer good enough and self-consistent mean field theory is expected to achieve a better result by promoting the bare $G^0$ to the dressed $G$ in the Hartree diagram. We have shown that by using this self-consistent prescription one can reproduce the DMFT result with very good agreement. Note that although this step restores self-consistency in terms of the density, the sum rules are no longer satisfied. Since this obvious shortcoming of TPSC in the partitioned lattice affects only the constant term, we assume that the description of the momentum-dependence is still accurate.

\end{document}